\documentclass[%
aps,prb,preprint,
superscriptaddress,
nobibnotes,
amsmath,amssymb,
floatfix,
]{revtex4-2}

\usepackage{xcolor}
\usepackage{graphicx}% Include figure files
\usepackage{dcolumn}% Align table columns on decimal point
\usepackage{bm}% bold math

\usepackage{caption}
\begin{document}

\title{Engineering the formation of spin-defects from first principles}

\author{Cunzhi Zhang}
\affiliation{Pritzker School of Molecular Engineering, University of Chicago, Chicago, IL
60637, USA}
\author{Francois Gygi}
\affiliation{Department of Computer Science, University of California Davis, Davis CA 95616, USA}
\author {Giulia Galli}
\email{gagalli@uchicago.edu}
\affiliation{Pritzker School of Molecular Engineering, University of Chicago, Chicago, IL
60637, USA}
\affiliation{Department of Chemistry, University of Chicago, Chicago, IL 60637, USA}
\affiliation{Materials Science Division and Center for Molecular Engineering, Argonne National Laboratory, Lemont, IL 60439, USA}

\date{\today}

%TC:ignore
\begin{abstract}

The full realization of spin qubits for quantum technologies relies on the ability to control and design the formation processes of spin defects in semiconductors and insulators. We present a computational protocol to investigate the synthesis of point-defects at the atomistic level, and we apply it to the study of a promising spin-qubit in silicon carbide, the divacancy (VV). Our strategy combines electronic structure calculations based on density functional theory and enhanced sampling techniques coupled with first principles molecular dynamics. We predict the optimal annealing temperatures for the formation of VVs at high temperature and show how to engineer the Fermi level of the material to optimize the defect's yield for several polytypes of silicon carbide. Our results are in excellent agreement with available experimental data and provide novel atomistic insights into point defect formation and annihilation processes as a function of temperature.

\end{abstract}

%TC:endignore

\maketitle
\clearpage

\section{Introduction}

Spin defects in wide bandgap semiconductors are promising platforms for several quantum technologies, including quantum photonics, and quantum sensing and communication \cite{wolfowicz2021NRM_SSSD,awschalom2018NP_SSSD}.
Among spin qubit hosts, in recent years  silicon carbide (SiC) has emerged as an ideal  material, due to mature growth, doping and fabrication techniques \cite{lohrmann2017RPP_SiC,castel2020JPP_SiC}, with qubits realized  with silicon vacancies (V$_{\rm Si}$) and nitrogen-vacancy pairs (NV), divacancies (V$_{\rm C}$V$_{\rm Si}$), and carbon antisite vacancies (CAV). The V$_{\rm C}$V$_{\rm Si}$ in SiC (which we denote as VV) has attracted particular interest, due to its optical addressability \cite{koehl2011NAT_optical}, a near-infrared spin-photon interface \cite{christle2017PRX_Spin_Photo}, long coherence times \cite{seo2016NC_coherence} and high-fidelity readout via spin-to-charge conversion \cite{anderson2022SA_SCC}. While numerous studies of defects in SiC have focused on their physical properties, much less is known about their formation processes, whose control is critical for the integration of semiconductors hosting spin qubits within electronic and optical devices \cite{wolfowicz2021NRM_SSSD, castel2020JPP_SiC,kraus2017NL_write, lohrmann2017RPP_SiC,castel2021MQT_control, toyli2010NL_aperture,luhmann2021PSS_control}. 

Defects in SiC are usually generated via implantation, irradiation or pulse laser, and by subsequent thermal annealing at high temperature \cite{wolfowicz2021NRM_SSSD,castel2020JPP_SiC}. Several experimental methods have been used to monitor defect formation, including electron paramagnetic resonance (EPR), photoluminescence  (PL), and deep-level transient spectroscopy ~\cite{luhmann2018JPAP_FLU_DIA,luhmann2019NC_coulomb,wolfo2017NC_ESR_1000,karsthof2020PRB_PL_1000,carls2010PRB_EPR_1000,bathen2019PRB_Vc}. Recent
progress has been reported in achieving spatial localization of defects~\cite{toyli2010NL_aperture}, as well as in controlling their charge state \cite{anderson2019SCI_device}, performance and yield~\cite{luhmann2019NC_coulomb,favaro2017NC_dop}.
% concentration by tunning ion or electron dose and defect orientation. 
In the case of the VV, one of the most studied defects in SiC, it is well established that $n$-doping conditions are beneficial to its formation 
\cite{karsthof2020PRB_PL_1000,wang2013JAP_VSi,bock2003PRB_VSi,yan2020JAP_VSi}, and
%under these conditions it is possible, for example, to prevent the conversion of Si vacancies into  immobile carbon antisite-vacancy  complexes (CAV) ~\cite{karsthof2020PRB_PL_1000,wang2013JAP_VSi,bock2003PRB_VSi,yan2020JAP_VSi}.
% Vsi migration ~ 750 C is inffered; the references provide weak evidences
a  lower bound for the annealing temperature ($T_{\rm Ann}$) required to generate VVs, $\sim$ 1,000 K, has been  estimated experimentally~\cite{karsthof2020PRB_PL_1000,son2006PRL_EPR_850,kobay2021JPAP_PL_850,dietz2022APL_Ann_850,son2021JAP_EPR_Fail}. However, different experiments have reported different  temperatures~\cite{son2006PRL_EPR_850,li2022NSR_Ann_900,chris2017PRX_Ann_745,falk2013NC_Ann_750_900,wolfo2017NC_ESR_1000,anderson2022SA_SCC,magnu2018PRB_Ann_800,chris2015NM_Ann_750,lin2021PRB_Ann_1050,son2021JAP_EPR_Fail,dietz2022APL_Ann_850}, 
with an optimal  $T_{\rm Ann}$ often quoted around 1,150 K~\cite{son2006PRL_EPR_850,ruhl2018APL_PL_900,almut2022APL_PL_800,kobay2021JPAP_PL_850}. The experimental determination of activation and optimal annealing temperatures remains a challenging task, because these quantities are usually inferred from the intensity of  EPR/PL signals which are affected by several factors, including the charge state and concentration of  defects~\cite{fuchs2015NC_PL_Fail}, Fermi-level position ($E_{\rm F}$)~\cite{wolfo2017NC_ESR_1000,son2021JAP_EPR_Fail}, and specific synthesis conditions~\cite{wang2019ACSP_PL_Fail,kasper2020PRA_PL_Fail}. Recently, the pairing of  V$_{\rm C}$ and V$_{\rm Si}$ into neutral VVs has been investigated theoretically, providing the first atomistic insight into the formation process~\cite{lee2021NC_VV}. 

However, as is the case for most point defects in semiconductors, our understanding of the VV formation mechanism at the atomistic level remains preliminary and qualitative. In particular, a relation between the host $E_{\rm F}$ and $T_{\rm Ann}$ has not yet been established, which is of great importance to control defects' formation, and an upper bound to $T_{\rm Ann}$ is yet unknown. Moreover, the dynamics of VV ~\cite{lee2021NC_VV,yan2020JAP_VSi},  the conditions for the defect immobilization in the lattice  and the effect of temperature on formation processes are only partially  understood.  
Addressing these open problems is difficult from an experimental standpoint, especially in the presence of   limited microscopic resolution, and atomistic simulations are key tools to gain detailed insights. 

Here we present a general computational protocol, based on first principles calculations, to study the formation of point defects in covalently bonded materials; in addition we provide specific predictions on optimal conditions for the formation of double vacancies in SiC.
We focus on the cubic phase (3C-SiC) for its simplicity with only one type of lattice site, and we discuss implications of our results for hexagonal polytypes.  We determine the preferred pathways leading to the VV formation and optimal values of $T_{\rm Ann}$ and $E_{\rm F}$,  and we elucidate the interdependence of these parameters. 
Our results point at the importance of considering multiple charge states of defects, as well as of configurations that are not thermodynamically stable, for accurate predictions of formation pathways. On the other hand, the sampling of paths with different spin states has a negligible impact on our predictions.

\section{Results}

\subsection{Computational strategy}

\begin{figure*}[!htp]
\centering
  \includegraphics[width=0.95\textwidth]{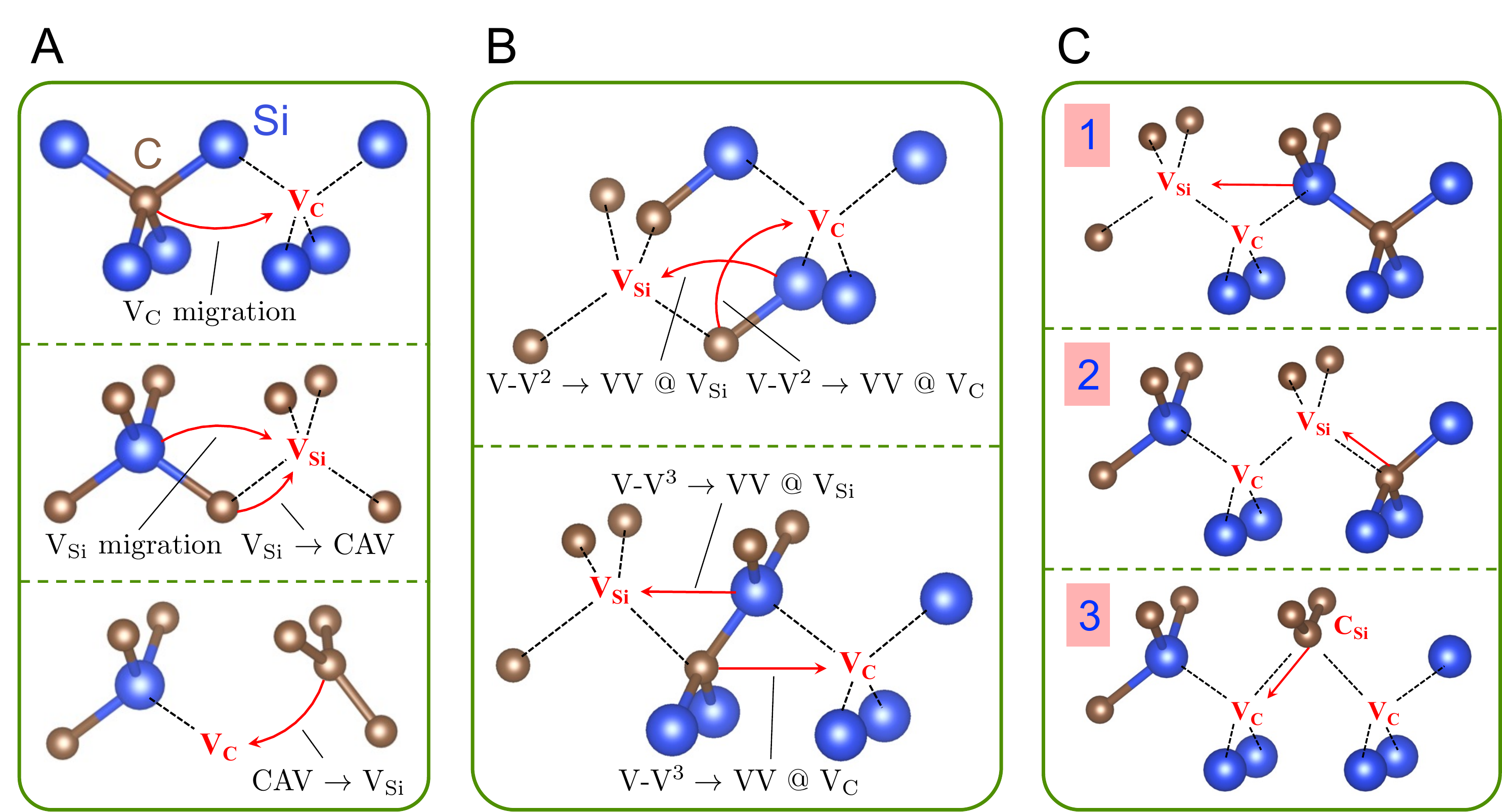}
  \caption{Atomic pathways investigated in this work. A: Monovacancy dynamics, including carbon (V$_{\rm C}$) and silicon (V$_{\rm Si}$) vacancy migration, and V$_{\rm Si}$ and carbon antisite vacancy complex (CAV) inter-conversion. B: Pairing of second (V-V$^2$) and third (V-V$^3$) neighbors V$_{\rm C}$ and V$_{\rm Si}$ vacancies to form a double vacancy VV. Only V-V up to third neighbors were considered, due to the size of our supercells. C: VV migration path with the lowest barrier, where several steps are illustrated. V$_{\rm C}$C$_{\rm Si}$V$_{\rm C}$ complex in 3 is denoted as VCV.}
  \label{fig:path}
\end{figure*}

{\it Pathways --} We studied defect dynamics and transformations during the thermal annealing process following defect generation by e.g. particle irradiation. We considered several possible processes relevant to the  formation of VVs in 3C-SiC, based on previous studies \cite{bock2003PRB_VSi,wang2013JAP_VSi,lee2021NC_VV,yan2020JAP_VSi}, and on our chemical intuition; they are summarized in Fig. \ref{fig:path}.  
 In addition to formation, we also considered dissociation processes, specifically  CAV $\rightarrow$ C$_{\rm Si}$ + V$_{\rm C}$ (where C$_{\rm Si}$ is an isolated carbon antisite) and  VV $\rightarrow$ V$_{\rm C}$ + V$_{\rm Si}$, which involve  multi-step migrations of mono-vacancies (MV) (not explicitly shown in Fig. \ref{fig:path}).

We did not consider interstitial (e.g. Si or C interstitial), and substitutional (e.g. N  substitution or C$_{\rm Si}$) defects; the former are expected to be annealed out once the paths described in Fig. \ref{fig:path} occur \cite{bock2003PRB_VSi,yan2020JAP_VSi}, and the latter are immobile at $\sim$ 1,000 K~\cite{kimoto2014book_SiC}.
  
% Their atomic pathways were not shown, because elementary steps during the dissociation, e.g. MV migration, are included in Fig. \ref{fig:path} {\bf unclear what you mean by this} 

\begin{figure*}[!htp]
\centering
  \includegraphics[width=0.7\textwidth]{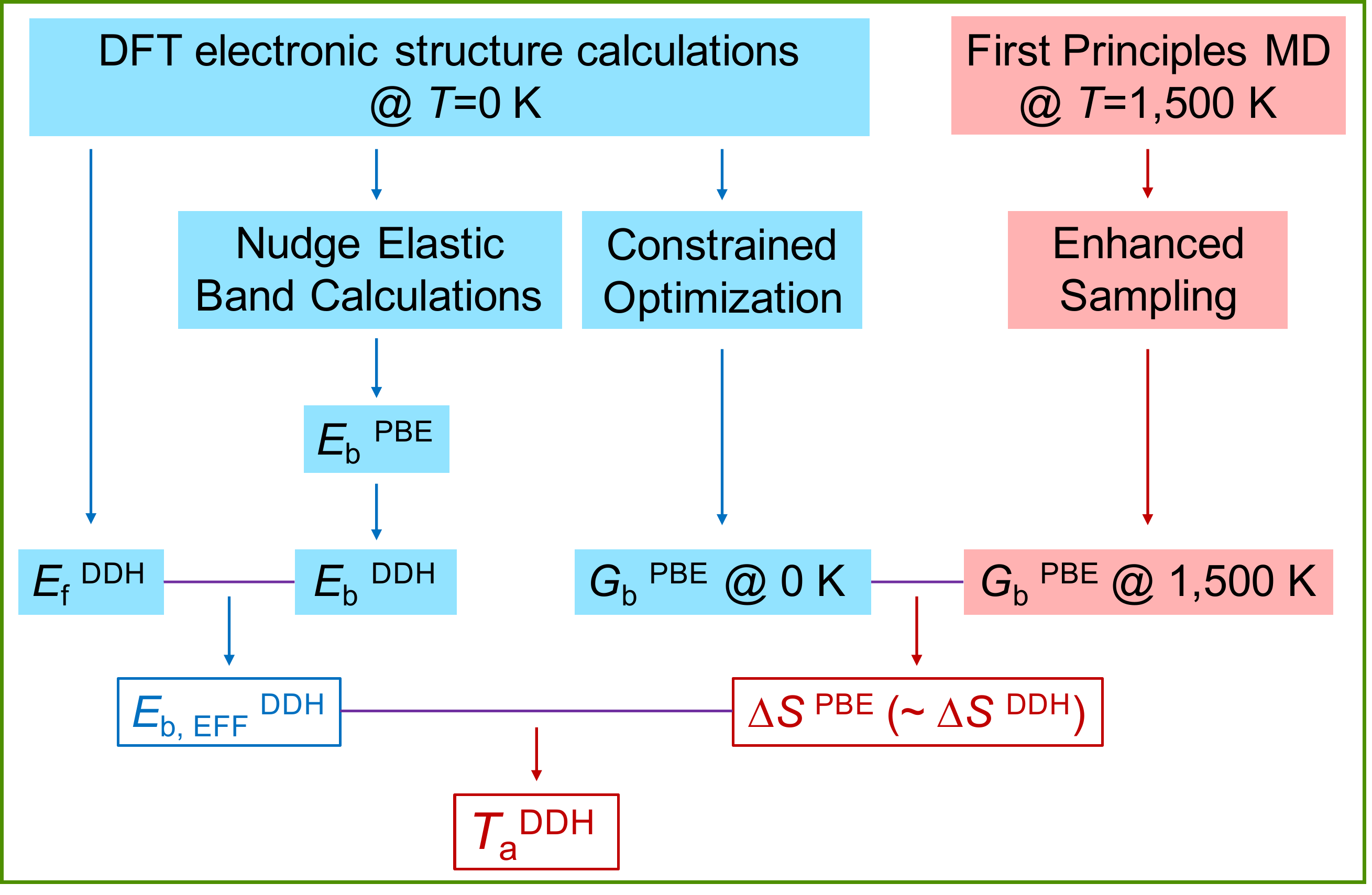}
  \caption{Computational Protocol. The calculations highlighted in blue (red) were carried out at zero (finite) temperature ($T$). We specify the functionals used in the calculations (PBE and DDH, see text) and the computed quantities: defect formation energies ($E_{\rm f}$), energy  and effective energy barriers ($E_{\rm b}$ and $E_{\rm b,\ EFF}$), Gibbs free energy bariers ($G_{\rm b}$), entropy differences ($\Delta S$) and activation temperature ($T_{\rm a}$). We obtained $E_{\rm b}^{\rm PBE}$ using the nudged elastic band method, then corrected our results using the DDH functional to obtain $E_{\rm b}^{\rm DDH}$ (see Methods).
  Assuming error cancellation, we considered  $\Delta S^{\rm PBE} \sim \Delta S^{\rm DDH}$ (see SI).}
  \label{fig:flowchart}
\end{figure*}

{\it Simulation protocol --} The simulation protocol used in our work is presented in Fig. \ref{fig:flowchart}. We studied the processes displayed in Fig. \ref{fig:path} using density functional theory (DFT) calculations with both the Perdew-Burke-Ernzerhof (PBE) and dielectric-dependent hybrid (DDH)  functionals, and we considered several charge ($q$) and spin ($s$) states (see Methods).
Specifically,  we considered different $s$ states to  determine the minimum energy path and energy barrier $E_{\rm b}$ as a function of $q$, and for a given pathway, we obtained an effective barrier, $E_{\rm b,\ EFF}$, as a function of the Fermi level $E_{\rm F}$:  
% Then, using our computed values of $E_{\rm f}$ and $E_{\rm b}$, we obtained an effective barrier $E_{\rm b,\ EFF}$ to characterize a given transition, as a function of $E_{\rm F}$:
\begin{equation}
\label{eq:eff_b}
   E_{\rm b,\ EFF} (E_{\rm F}) = \underset{q}{\text{min}}
   \big\{ \Delta E_{\rm f}(q, E_{\rm F})
   + E_{\rm b}(q) \big\}
\end{equation}
where $\Delta E_{\rm f}(q, E_{\rm F}) = E_{\rm f}(q, E_{\rm F}) - \underset{q}{\text{min}}\{ E_{\rm f}(q, E_{\rm F}) \} $ is the formation energy difference relative to the most stable charge state, for a specific value of $E_{\rm F}$, which in Eq. \ref{eq:eff_b} is treated as a parameter; $E_{\rm f}$ is the formation energy of a defect in the initial state and $E_{\rm b}$ denotes barriers between the initial and transition state. Note that $E_{\rm b,\ EFF}$ is a continuous function of $E_{\rm F}$, while $E_{\rm b}$ exhibits steps at charge transition levels.
The expression of $E_{\rm b,\ EFF}$ in Eq. \ref{eq:eff_b} assumes that charge state equilibration processes are  faster than the transformation of defects into different configurations. We verified the validity of this assumption  at high $T$ ($\sim$  1,000 K; see SI).
% Here, we validated the charge state equilibration should be faster than defect transformations, e.g. at $\sim$  1000 K. 
% which is justified by the fast $q$ transition at high $T$. 
We emphasize that thermodynamically unstable $q$ states may participate and play an important role in  defect transformation processes, since exploring those states may lead to lower effective barriers.

As mentioned above, the Fermi level is a parameter in Eq. \ref{eq:eff_b}, and we estimated the experimental conditions that may lead  to specific, desired values of $E_{\rm F}$ based on charge neutrality conditions and the electronic properties of the system  (see SI). 

We estimated the entropy change $\Delta S$ from the initial to the transition state by computing the difference in free energy barriers $\Delta G_{\rm b}$ between 0 and 1,500 K, where  $G$ at 0 K is:
\begin{equation}
  G(\xi^0, 0 {\rm K}) = 
  \underset{ \textbf{\textit {x}} }{\text{min}}\ 
  U(\textbf{\textit {x}}) 
  \big|_{ \xi(\textbf{\textit {x}}) = \xi^0 }
\end{equation}
and $\xi$ is a collective variable (the choice of collective variables is described in the Methods and SI); $U$ is the potential energy and $\textbf{\textit {x}}$ are atomic coordinates.
We calculated $G$ at high temperature, specifically  $T=1,500$ K, using first-principles molecular dynamics (MD) and the adaptive biasing force method (see Methods), and 
 we estimated $\Delta S$  as $\Delta G_{\rm b} / T$ (see SI). 
 Due to the computational cost, we obtained $\Delta S$ for only three paths (see SI). 

Once we obtained $E_{\rm b,\ EFF}$ and $\Delta S$, we could compute the temperature $T_{\rm a}$, above which a given process is thermally activated, and for which we used the harmonic transition state theory:
\begin{equation}
\label{eq:ta}
  T_{\rm a} =  \Big[ k_{\rm B} 
  \ln( \Gamma_{0} \exp( \frac{\Delta S}{k_{\rm B}}) 
  / \Gamma ) \Big]^{-1} 
  \times E_{\rm b,\ EFF}
\end{equation} 
where $\Gamma_{0}$ denotes an attempt frequency and $\Gamma$ a jump frequency. The values chosen for $\Gamma_{0}$, $\Gamma$ and $\Delta S$  are given in the SI. A simple sensitivity analysis, also in the SI,  shows that in Eq. 3 the prefactor is relatively insensitive to the choice of these values. In addition, we systematically investigated the effect of thermal expansion and that of entropy  on  computed activation temperatures, amounting to variations in  $T_{\rm a}$ of less than 10 $\%$ (see SI).

% computed at 1,500 K are lower by $\sim$ by (0.12, 0.38) (10 per cent of the Barrier) eV than those computed at 0K (see SI), consistent with estimates based on the harmonic approximation\cite{rauls2003PRB_TS}. 

\subsection{Theoretical predictions}

\begin{figure*}[!htp]
\centering
  \includegraphics[width=0.88\textwidth]{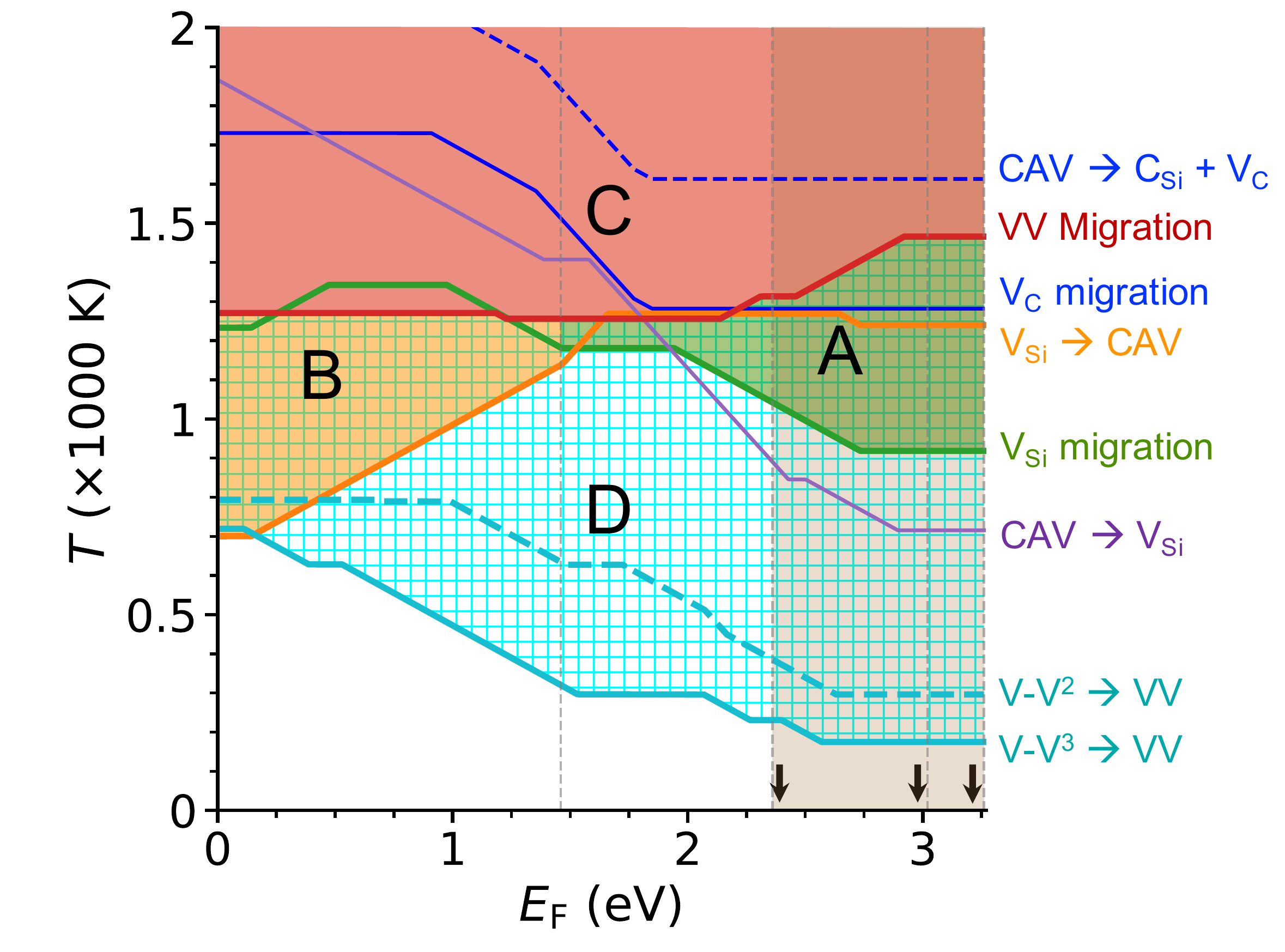}
  \caption{ Computed activation temperature
  $T_{\rm a}$ as a function of the Fermi level $E_{\rm F}$ (referred to the top of the valence band). The processes considered in this work are shown on the right hand side of the figure and they are summarized in Fig. \ref{fig:path}; the notation V-V$^{2/3} \rightarrow$ VV refers to V-V$^{2/3} \rightarrow$ VV @ V$_{\rm Si}$ (Fig. \ref{fig:path}B), a path  with lower barrier than V-V$^{2/3} \rightarrow$ VV @ V$_{\rm C}$ . The lines indicate the temperature, as a function of $E_{\rm F}$, above which the process indicated on the right hand side is activated. Regions A, B, C and D where specific processes occur are described in the text. The arrows indicate the conduction band minimum of  the various polytypes, which was computed by aligning their respective valence band maxima; they are shown in increasing order of energy for 3C, 6H and 4H of SiC. 
  % the polytypes 3C, 6H and 4H of SiC, in increasing order of energy, respectively.
  }
  \label{fig:phase}
\end{figure*}

We start by presenting our results for 3C-SiC  and we  report our predictions for the activation temperature $T_{\rm a}$ for various processes.

% In Fig. \ref{fig:phase}, we report our results for the activation temperature $T_{\rm a}$ as a function of $E_F$, computed for multiple processes. In addition we estimate the appropriate $T_{\rm Ann}$ and $E_{\rm F}$ for various processes to occur, to guide experiments achieve a  better control of defect formation. We first discuss our results for 3C-SiC  ($E_{\rm F} <$ 3C-CBM) and then  the implications of our findings for hex-SiC. 
{\it Activation Temperature --} In Fig.\ref{fig:phase} we show  $T_{\rm a}$ as a function of $E_{\rm F}$, where  lines indicate the values above which a given process can occur.
%Note the strong dependence of  $E_{\rm b,\ EFF}$ (and hence $T_{\rm a}$) 
%on $E_{\rm F}$, consistent with experiments showing that the charge states of point defects (whose stability depends on $E_{\rm F}$) drastically influence the formation kinetics\cite{luhmann2021PSS_control,favaro2017NC_dop,luhmann2019NC_coulomb}.  
We find that for all values of the Fermi level, the onset of V$_{\rm Si}$ migration occurs at temperatures lower than those activating  V$_{\rm C}$ diffusion, consistent with the results of previous studies \cite{karsthof2020PRB_PL_1000,zolnai2004JAP_Vc_1100,dietz2022APL_Ann_850,almut2022APL_PL_800,ruhl2018APL_PL_900,kobay2021JPAP_PL_850,carls2010PRB_EPR_1000,defo2018PRB_VSi,wang2013JAP_VSi,bock2003PRB_VSi,yan2020JAP_VSi}. Above 1,000 K, V$_{\rm Si}$ can diffuse, and hence when migrating  it may lead to the formation of VV. 
Interestingly, our calculations show that the pairing of mono-vacancies  is facilitated by the Coulomb interaction between V$_{\rm Si}^{\ -}$ and V$_{\rm C}^{\ +1/+2}$. Indeed, we find that for $1.46 < E_{\rm F} < 1.85 $ eV,  V$_{\rm Si}^{\ -}$ and V$_{\rm C}^{\ +1/+2}$ are respectively the most stable charge states of the two mono-vacancies (whether V$_{\rm C}$ is in charge state +1 or +2 depends again on the Fermi level).
It is important to note that for $E_{\rm F}$ $<$ 1.85 eV, a simple consideration based on energy barriers $E_{\rm b}$ would yield  $T_{\rm a}$ $\sim$ 1,500 (1,700) K as the temperature required for a carbon vacancy to migrate in a stable charge state $q$ = +1 (+2). However, upon computing {\it effective} barriers, we find a process with lower  $T_{\rm a}$ (as low as $\sim$ 1,300 K); such process   involves intermediate charge states that are not thermodynamically stable but nevertheless allows for paths with lower barriers. Specifically, we find that thermal vibrations and changes in carrier density at high $T$ can cause a transition from  V$_{\rm C}^{\ +1/+2}$ to V$_{\rm C}^{\ 0/+1}$ charge states, and that the latter migrate through a path with a barrier lower than that of  V$_{\rm C}^{\ +1/+2}$, before returning to the original charge state (see SI).

The migration of V$_{\rm Si}$ discussed above is a necessary but not a sufficient condition for the formation of VV. We find, in agreement with previous studies \cite{karsthof2020PRB_PL_1000,wang2013JAP_VSi,bock2003PRB_VSi,yan2020JAP_VSi}, that it is important to realize, at the same time, $n$-type conditions. In particular $E_{\rm F}$ should be above 1.46 eV (see Fig. \ref{fig:phase}).
Indeed, under $p$-type conditions ($E_{\rm F} <$ 1.46 eV), the V$_{\rm Si}$ $\rightarrow$ CAV process is energetically favored over monovacancies diffusion (region B in Fig. \ref{fig:phase})  and V$_{\rm Si}$ is trapped into the CAV complex and becomes immobile \cite{bock2003PRB_VSi,wang2013JAP_VSi}. Once formed, 
CAV remains stable as both the back conversion to V$_{\rm Si}$ (CAV $\rightarrow$ V$_{\rm Si}$) and its dissociation (CAV $\rightarrow$ C$_{\rm Si}$ + V$_{\rm C}$) are unlikely below 1,500 K due to high free energy barriers.
% CAV does not diffuse below $ xx$ K and the back conversion V$_{\rm Si}$ (CAV $\rightarrow$ V$_{\rm Si}$) is unlikely due to a high free energy barrier. 
Instead, under $n$-type conditions ($E_{\rm F} > $ 1.46 eV) the migration of V$_{\rm Si}$ is an energetically favored process and  VV creation may occur
% ({\bf explain why; is it because different doping conditions give rise to different barriers?}) 
(region A in Fig. \ref{fig:phase}). We note that  in general, the higher $E_{\rm F}$, the more favorable the conditions for VV formation for several reasons. Increasing  $E_{\rm F}$ leads to a lower MV migration barrier and increased mobility for V$_{\rm Si}$, and to  a higher (reduced) barrier for V$_{\rm Si}$ $\rightarrow$ CAV (CAV $\rightarrow$ V$_{\rm Si}$), leading to  a lower probability of  CAV formation and a higher probability of  CAV conversion to V$_{\rm Si}$.
 
% The generation of V$_{\rm C}$ out of V$_{\rm Si}$, via V$_{\rm Si}$ $\rightarrow$ CAV $\rightarrow$ C$_{\rm Si}$ + V$_{\rm C}$, is blocked unless at $T >$ 1600 K.
% \textcolor{blue}{former mentioning of CAV dissociation? [supply]}

An additional condition for the formation of VVs is an annealing $T$ below 1,300 K. We find that VV can migrate  for  $ T >\ \sim$ 1,300 K and will likely either  form large complexes (e.g. VV + V$_{\rm C}$) \cite{bock2008PSS_cluster} or diffuse and  eventually move  to the surface of the sample (region C in Fig. \ref{fig:phase}). These processes undermine the stability and abundance of double vacancies.
In addition we find that V$_{\rm C}$ is immobile below 1,300 K, which  is overall a favorable condition for VV formation. Indeed we expect V$_{\rm C}$ to be abundant in experimental samples, due e.g. to a formation energy lower than that of V$_{\rm Si}$ and other point defects, and it can be incorporated in a VV $+$ V$_{\rm C}$ cluster if it migrates. Therefore, we suggest that $T_{\rm Ann}$ should be $<$ 1,300 K for optimal yield, stability and localization of VV defects. 

Note that larger vacancy clusters can be formed also by incorporating V$_{\rm Si}$ \cite{wang2013JAP_VSi} into VV, and this  undesirable process can only  be mitigated by reducing the concentration of V$_{\rm Si}$. 
Unfortunately, once VV is formed, charge-state engineering \cite{luhmann2021PSS_control,favaro2017NC_dop,luhmann2019NC_coulomb} is not an  effective tool to hinder the formation of larger vacancy clusters because the most stable state of VV is neutral for $E_{\rm F}$ above mid-gap. 
% Hence doping engineering may not be used to influence the separation of VV from monovancies, that are stable in charged states.

Other defects of potential concern for the stability and formation of the VV are single interstitials (C$_{\rm i}$ and Si$_{\rm i}$). For example, C$_{\rm i}$ could be re-emitted from C$_{\rm i}$ clusters at high $T$, and subsequently aggregate with VV. Fortunately, C$_{\rm i}$ emission is unlikely to occur below 1,300 K, according  to previous DFT calculations  (barrier $>$ $\sim$ 4 eV) \cite{bock2008PSS_cluster,bock2004PRB_cluster,matta2004PRB_cluster}. Nonetheless, VV could be annihilated by the presence of C$_{\rm i}$ if some weakly bonded C$_{\rm i}$ clusters turn out to be present in the sample.
% , dissociation plus migration energy, is mostly $>$ 4 eV in SiC\cite{bock2008PSS_cluster,bock2004PRB_cluster}, translated into $T_{\rm a} >\ \sim$ 1320 K.

\begin{figure*}[!htp]
\centering
  \includegraphics[width=0.99\textwidth]{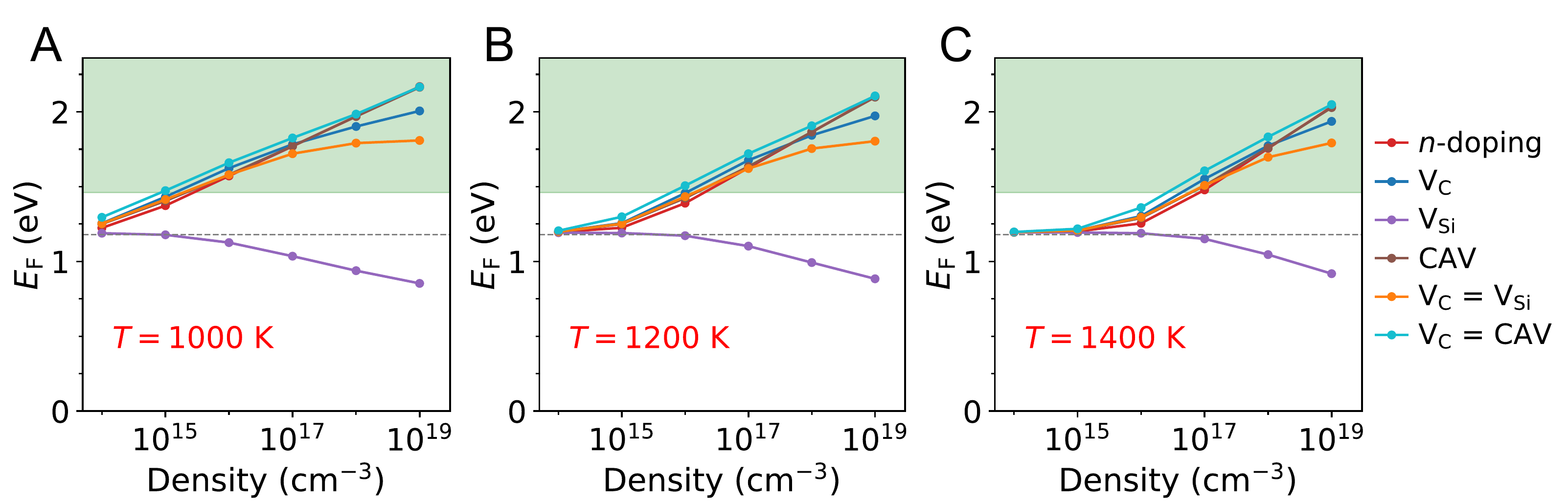}
  \caption{Fermi level ($E_F$) as a function of defects density. The Fermi level is referred to the top of the valence band. We show results for different temperatures (1,000 K (A), 1,200 K (B) and 1,400 K (C)); we consider, separately, initial $n$-doping of the sample, and specific concentrations of carbon (V$_{\rm C}$) and silicon (V$_{\rm Si}$) vacancies,  and carbon antisite vacancy complexes (CAV); we also consider two additional cases: (i) same concentration of V$_{\rm C}$ and V$_{\rm Si}$ (V$_{\rm C}$ $=$ V$_{\rm Si}$); (ii) same concentration of V$_{\rm C}$ and CAV (V$_{\rm C}$ $=$ CAV).
  The dashed line indicates the value of  mid-gap for 3C-SiC; the green-region for $E_{\rm F}$ $>$ 1.46 eV indicates favorable conditions for the formation of the double vacancy in the range of temperatures between 1,000 and 1,300 K (see text).}
  % % the solubility limit for major dopants, N, P, Al, B, in SiC $\sim$ 1e20 cm-3 [kimoto book]
  % % --> A criteria to tell if a doping density is high or low
  \label{fig:fermi}
\end{figure*}

To estimate the optimal annealing $T$ in the range of (1,000, 1,300) K, we consider the dependence of the Fermi level on doping densities. In 3C-SiC, maintaining $E_{\rm F} >$ 2 eV requires a rather large doping density $>$ $\sim$ 10$^{18}$ cm$^{-3}$ (see Fig. \ref{fig:fermi}). Therefore, it is conceivable that a desirable Fermi level range is  1.46 $< E_{\rm F} <$ 2 eV, over which $T_{\rm a}$ for V$_{\rm Si}$ migration is a constant, roughly equal to 1,200 K.  For  $T_{\rm Ann} $ between 1,200-1,300 K the Fermi level would be lower, for a fixed doping density, than in the range 1,000-1,200 K, hence possibly leading  to  V$_{\rm Si}$ trapping at CAV defects. Therefore we conclude that the optimal $T_{\rm Ann} \sim$ 1,200 K.

So far, we have identified {\it a} suitable range of $T_{\rm Ann}$ under $n$-type conditions, (1,000, 1,300) K, with an optimal value of 1,200 K. However, as shown in Fig. 3 (region D), there exist conditions at which  VV may form  also below 1,000 K,  as long as there are  V-V defects present in the sample.  Surprisingly, barriers are lower for the pairing of third than second neighbors' vacancies.
These results may help understand the conditions required for VV formation in small SiC nanoparticles (with diameter less than 10 nm), observed at lower $T_{\rm Ann}$, e.g. $\sim$  440 K \cite{beke2020JPCL_NP}, than in the bulk, since in nanoparticles MV separation distances are usually smaller.
 % These results indicate that VV formation in nano-particles should occur  at lower $T_{\rm Ann}$ than in the bulk, e.g. $\sim$  140 $^{\circ}$C as reported in Ref. \cite{beke2020JPCL_NP}, since the MV separation distance is smaller in NP than in the bulk. For example in the bulk, the average spacing between MV can be estimated to be $\sim$ 20 \AA at  a density of $\sim$ 10$^{20}$ cm$^{-3}$.

{\it Fermi level and defect density--} We now turn to  exploring how  conditions identified above for VV formation may be achieved experimentally, by controlling for example the Fermi level and density of defects. 
In addition to the electronic properties of the system, $E_{\rm F}$  depends on $T$, initial sample doping and of course defect density (see SI) which, at each given time of the annealing process is the most elusive parameter. The spatial distribution of defect density may be non-uniform and  it depends  on specific dose and energy of particles used during the bombardment of the sample \cite{kasper2020PRA_PL_Fail,wolfowicz2021NRM_SSSD,castel2020JPP_SiC,luhmann2021PSS_control}.
In spite of these uncertanties, it is interesting to obtain a qualitative estimate of the doping conditions necessary to achieve the desirable Fermi level values for the formation of VVs. 
 In Fig. \ref{fig:fermi}, $E_{\rm F}$ is calculated at several $T$ and for various doping and defect densities. We find that the presence of V$_{\rm C}$ and/or CAV would induce $n$-doping while the presence of V$_{\rm Si}$ would induce $p$-doping in the sample. 
Hence, the required condition to reach $E_{\rm F}$ $>$ 1.46 eV within (1,000, 1,300) K,  is that at least one of the following concentrations--  $n$-doping (e.g. [N]),  C vacancies [V$_{\rm C}$] or antisite [CAV]--  be larger than 10$^{16}$ cm$^{-3}$.  

Note that $p$-doping  conditions induced by the presence of V$_{\rm Si}$, which are unfavorable for the formation of VVs,  may be  compensated  by the presence of V$_{\rm C}$  or CAV of comparable amounts (see  Fig. \ref{fig:fermi}).
Further, the V$_{\rm Si}$ to CAV conversion process, although it renders V$_{\rm Si}$ less mobile, helps  increasing $E_{\rm F}$ which in turns facilitates VV formation. These results emphasize the complex, interdependent role of multiple defects in tuning $E_{\rm F}$ and ultimately leading to the formation of VVs.
  
We conclude this section by discussing  VV formation properties in hexagonal (hex) polytypes, e.g., 4H-SiC. The extension of our results for 3C-SiC (where only $k$-sites are present)  to hexagonal lattices (where both $h$- and $k$-sites are present) should be considered as a qualitative prediction. In hexagonal samples, the variation in stability and barriers of defects occupying different lattice sites is negligible, compared to the  energy scale of $\sim$ several eV of most barriers computed in our calculations  \cite{yan2020JAP_VSi,karsthof2020PRB_PL_1000,wang2013JAP_VSi,defo2018PRB_VSi,bathen2019PRB_Vc}.
The position of the valence band maximum (VBM) is nearly the same in cubic and hexagonal SiC, while that of the conduction band minimum (CBM) is higher in hex-SiC \cite{luning1999PRB_CBM} (see Fig. \ref{fig:phase}). We find that  the creation of VVs  is more facile in hex-SiC; indeed the conditions of regions A (corresponding to formation and stability of VV) and D (corresponding to pairing of nearby vacancies) can be obtained in a slightly wider range of temperatures than in 3C-SiC (for values of the Fermi level attainable in hex-SiC) and, importantly, for lower doping densities. 
For example, in  4H-SiC,  under intrinsic condition, $E_{\rm F} \sim 1.6$ eV is larger than $1.46$ eV, and with a moderate $n$-doping $>$ 10$^{15}$ cm$^{-3}$, it may be increased above 2.0 eV at 1,200 K (see SI). 
We predict that an appropriate  $T_{\rm Ann}$ for hex-SiC is in the range of (900, 1,300) K, with an optimal value around 
  1,200 K.
% which should be sample dependent. 
% Besides, like V$_{\rm C}$\cite{bathen2019PRB_Vc}, we anticipate the anisotropy, e.g., in the V$_{\rm Si}$ and VV diffusion in hex-SiC. [Next paper]

Our predictions are in excellent agreement with several experimental observations. To synthesize VV, most experiments adopted $T_{\rm Ann}$ in a range of (1,050, 1,350) K,  consistent with our prediction of annealing $T$ of (1,000, 1,300) K in 3C-SiC and (900, 1,300) K in hex-SiC. 
Experimentally, the optimal $T_{\rm Ann}$ was determined by  PL or EPR maximum intensities  and found to be   $\sim$ 1,150 K, in agreement with our calculations of $\sim$ 1,200 K.  We emphasize that, depending on the experimental setup, the decrease in signal above $\sim$ 1,150 K is not necessarily related only to changes in VV concentration. 
We predict VV can be stable up to 1,300 K, above which its density decreases due to diffusion. This is consistent with the 
significant drop of VV signals in experiments as $T$ $>$ 1,300 K \cite{wolfo2017NC_ESR_1000}, and with the highest PL and EPR intensities detected at 1,300 K \cite{karsthof2020PRB_PL_1000,carls2010PRB_EPR_1000}.

\section{Discussion}

By combining DFT calculations with semilocal and hybrid functionals, nudged elastic band  and first principles MD simulations, we obtained  a detailed, atomistic description of the VV formation process in 3C-SiC. 
We computed energy barriers and activation temperatures for multiple defects and pathways as a function of the Fermi level $E_{\rm F}$. We then identified favorable conditions for the formation of VVs and discussed how suitable values of $E_{\rm F}$ can be obtained via careful tuning of doping or defects densities. Our calculations show that one should use $n$-doped samples with $E_{\rm F}$ $>$ 1.46 eV during annealing, to ensure the stability of single V$_{\rm Si}$, and $T_{\rm Ann}$ $>\ \sim$ 1,000 K to activate V$_{\rm Si}$ migration for aggregation with V$_{\rm C}$. Further,  $T_{\rm Ann}$ should be lower than  $\sim$ 1,300 K to suppress VV diffusion, thus ensuring its stability and immobilization, with the optimal  $T_{\rm Ann}$ estimated to be $\sim$ 1,200 K.
However, VV can also be created at lower $T$ from neighboring V$_{\rm C}$-V$_{\rm Si}$ pairs; these  may be present after irradiation or implantation, and may be prominent in SiC nanostructures, suggesting that the formation of VVs in small nanoparticles should occur at lower $T$ than in the bulk.
Our findings also suggest that VV signals may be detected  at low annealing temperatures, which however should not be interpreted as lower bounds for V$_{\rm Si}$ diffusion.
Moreover, we predict that VV formation in hex-SiC can be more facile than in 3C, due to a larger band gap and higher CBM position, which allow for the use of lower doping densities and lead to a slightly broader range of favorable annealing $T$.
Our results are in excellent agreement  with experiments, while providing new and improved understanding of formation mechanisms at the atomistic level. The knowledge obtained here may benefit the controlled fabrication and device integration of VV, assisting its applications for quantum technologies.

%Our investigation of VV is also informative for other qubits, such as V$_{\rm Si}$ and NV centers. 
%For example, the formation of the NV center in SiC is expected to be facilitated by the migration of  V$_{\rm Si}$ in nitrogen doped samples\cite{zarg2016PRB_NV,von2016PRB_NV}. Consequently, favorable $E_{\rm F}$-$T$ conditions identified here for VV formation and immobilization should be partially applicable to the NV center as well.

Importantly, the computational protocol and strategies developed here, based on first principles calculations, are general and can be readily extended to investigate defects in other covalently bonded materials. Multiple paths with different charge  states should be considered to understand point defect formation processes, taking into account thermodynamically unstable ones, which may facilitate the exploration of low barrier paths at high $T$.  Our findings show that it is key to conduct calculations of effective barriers as a function of the Fermi level, which itself depends on $T$, and not only of barriers between thermodynamically stable states. In addition, it is critical to consider not only formation but also annihilation pathways to obtain faithful predictions of formation processes. Unexpectedly, we found that although important for accurate quantitative predictions, 
thermal expansion and entropic contributions are not critical to determine general trends of activation temperatures for different paths. 

One important problem that remains to be addressed is the  influence, on defects' formation, of the specific synthesis procedures, e.g., by irradiation of the sample.  
Using our  computed energy barriers as input, one can simulate real-time defect evolution, e.g., via kinetic Monte Carlo methods, which could then provide information about optimal annealing times.
These possible directions are worthy of future explorations.

%TC:ignore
\section{Methods}

\subsection{Density functional theory calculations}
We performed DFT calculations using the Qbox \cite{gygi2008_qbox} and the Quantum Espresso \cite{giann2009JPCM_QE} codes. We used the PBE \cite{perdew1996PRL_GGA} and DDH \cite{skone2014PRB_DDH} (15\% exact exchange) functionals, optimized norm-conserving Vanderbilt pseudo-potentials \cite{schlipf2015CPC_ONCV}, a plane-wave kinetic energy cutoff of 60 Ry. We conducted calculations in 216 atom supercells with lattice constant 4.416 Å, and with either the $\Gamma$ point or a 2$\times$2$\times$2 Monkhorst-Pack grid to sample the Brillouin zone. The lattice constant was determined by first-principles MD (FPMD) simulations in the NPT ensemble at 1,500 K at the PBE level of theory.  We considered structural relaxations as converged when  residual forces on atoms were  $<$ 0.01 eV/\AA. We considered charge state $q$ from -2 to 2 for all defects, expect for V$_{\rm C}$  where $q$ = 0, 1 or 2; spin state $s$ = [S, T] ([D, Q]) for even (odd) number of electrons, where S: singlet, D: doublet, T: triplet, Q: quartet. 
We chose not to employ empirical force-fields, which would have allowed for the use of larger supercells, as they are not appropriate to simulate $q$ and $s$ degrees of freedom; in addition we found that in several cases many of the popular force-fields used for SiC  cannot reproduce  DFT results. 
\subsection{Nudged elastic band calculations}
We carried out climb image nudged elastic band (CI-NEB) simulations at the PBE level with 2$\times$2$\times$2 k-point grids, by coupling Qbox with the  PASTA \cite{kundu2018CPC_PASTA} code. We used spring constants of 2 eV/\AA$^2$ and force tolerance of 0.02 eV/\AA. We determined the most stable spin state among [S, T] or [D, Q] for each NEB image at a given charge state $q$; the corresponding total energies and atomic forces were then used to update NEB images to determine the minimum energy path and energy barriers $E_{\rm b}$. In this way, $E_{\rm b}$ is only a function of $q$. For most pathways studied here, the most stable spin state remain the same along the whole path; for those paths for which we observed a change of spin states, we found that the energy splitting between different spin states at the transition-state is generally small, i.e. less than  10 $\%$ of $E_{\rm b}$. 
% We found that the energy splitting between different spin states was always small, compared to energy   barriers of $\sim$ several eV. 

We then computed total energies for converged images, at the PBE and DDH level of theory using only the $\Gamma$ point \cite{bathen2019PRB_Vc}. We denote the barriers obtained in this way as $E_{\rm b}^{\rm PBE}$ @ $\Gamma$ and $E_{\rm b}^{\rm DDH}$ @ $\Gamma$. We computed  the correction  to apply to PBE results in order to estimate DDH barriers as:  [$E_{\rm b}^{\rm DDH}$ @ $\Gamma - E_{\rm b}^{\rm PBE}$ @ $\Gamma$]. We added such correction to  $E_{\rm b}^{\rm PBE}$ @ 222 (barriers computed with the 2$\times$2$\times$2 k-point grid)   to obtain $E_{\rm b}^{\rm DDH}$ @ 222. Here, we assumed that the minimum energy paths at the PBE and DDH level of theory are similar; energy difference calculated with the $\Gamma$ point differed only slightly from those obtained with the  2$\times$2$\times$2 $k$-mesh. The results reported in the main text were obtained with $E_{\rm b}^{\rm DDH}$ @ 222.

\subsection{Formation energy calculations}
The formation energy of defect X in charge state $q$, $E_{\rm f} ( {\rm X}^q )$, was computed as:

\begin{equation}
   E_{\rm f} ( {\rm X}^q ) = 
   E_{\rm tot}( {\rm X}^q ) - E_{\rm tot}( {\rm SiC} ) 
   - n_{\rm C}\mu_{\rm C} - n_{\rm Si}\mu_{\rm Si} 
   + q E_{\rm F} + E_{\rm corr}( {\rm X}^q )
\end{equation}
where $E_{\rm tot}( {\rm X}^q )$ is the total energy of a SiC supercell with ${\rm X}^q$; $E_{\rm tot}( {\rm SiC} )$is the total energy of the pristine SiC supercell; $\mu_{\rm C}$ and $\mu_{\rm Si}$ are chemical potential of C and Si; $n_{\rm C}$ and $n_{\rm Si}$ are number of added ($+$) or removed ($-$) C or Si atoms to form X, respectively; $E_{\rm F}$ is the Fermi energy referred to the VBM; $E_{\rm corr}( {\rm X}^q )$ is the energy correction for spurious electrostatic interactions present in supercell calculations.

Using  relaxed configurations at the PBE level of theory, we computed the total energy and electrostatic potential using Quantum Espresso and the DDH functional. We obtained $E_{\rm corr}$ using the  method developed by Freysoldt, Neugebauer, and Van de Walle \cite{FNV2009PRL_correction}. We used a dielectric constant equal to 9.72. The chemical potential $\mu_{\rm C}$ was calculated as the energy per atom in diamond; $\mu_{\rm Si}$ was calculated as $\mu_{\rm SiC} - \mu_{\rm C}$, where $\mu_{\rm SiC}$ is the energy per formula unit in bulk SiC. The results are shown in Fig. S1.
% of Si and C are calculated as energy per atom in bulk Silicon and diamond.  C-rich condition was adopted in this work. 
 
Finally, 
% we estimated the C$_{\rm Si}$ and V$_{\rm C}$ 
binding energies between defects are required to compute the barriers of the CAV/VV dissociation processes. We estimated the C$_{\rm Si}$ and V$_{\rm C}$ binding energies as $\sim$ 1 eV from previous studies \cite{bock2003PRB_VSi,wang2013JAP_VSi}. We directly computed the V$_{\rm C}$ and V$_{\rm Si}$ binding energies, which are $\sim$ 3 eV for $E_{\rm F}$ near the mid-gap of 3C-SiC.  
% 1) CAV dissociation, i.e., CAV $\rightarrow$ C$_{\rm Si}$ + V$_{\rm C}$; 2) VV dissociation, i.e., VV $\rightarrow$ V$_{\rm C}$ + V$_{\rm Si}$. 

\subsection{Enhanced sampling calculations}
We computed free energies of  defect transformations by coupling the Qbox and SSAGES~\cite{sidky2018JCP_SSAGES} codes. We used Qbox to perform FPMD in the NVT ensemble and the adaptive biasing force method~\cite{darve2008JCP_ABF} in SSAGES to calculate free energy gradients. We utilized the collective variable (CV) $\xi_{\rm C/Si}$ \cite{bock2003PRB_VSi}:
\begin{equation}
   \xi_{\rm C/Si} = \Big( \bm{R}_{\rm C/Si} - \frac{1}{M} 
   \sum_{i \in \rm gate\ atoms} m_{i} \cdot \bm{R}_{i} \Big)
   \cdot \bm{e}_{\rm projection}
\end{equation}
 
where $\bm{R}_{\rm C/Si}$ are the coordinates of moving C/Si atoms; $m_i$ is the  mass of the $i^{\rm th}$ gate atom; $\bm{R}_{i}$ is the coordinate of the $i^{\rm th}$ gate atom; $M$ is the total mass of the gate atoms; $\bm{e}_{\rm projection}$ is the  unit projection vector (see Fig. S2). 

We carried out free energy calculations for the three processes presented in Fig. S2, where the definition of CVs and gate atoms is specified. For each path, we performed FPMD simulation at 1,500 K using a time step of 1 fs, for $\sim$ 370 ps. For computational efficiency, we used the PBE functional, 40 Ry kinetic energy cutoff and the $\Gamma$ point; we considered defects only at $q$ = 0 and $s$ = T in our MD simulations.
% We obtained the free energy barrier, and named it as $G_{\rm b}^{\rm PBE}$ @ 1500.

To elucidate the effect of $T$ on barriers, we also computed free energy profiles at 0 K. We used the  same functional and cutoff as in FPMD simulations for consistency; we used the Sequential Least SQuares Programming method \cite{kraft1988_SLSQP}  in the SciPy package to carry out constrained optimizations along one-dimensional CVs. 
% Accordingly, we named the free energy barrier as $G_{\rm b}^{\rm PBE}$ @ 0. Based on harmonic approximation and classical statistic, we computed $\Delta S$ as: $($ $G_{\rm b}^{\rm PBE}$ @ 0 $-$ $G_{\rm b}^{\rm PBE}$ @ 1500 $)$ $/$ 1500 $\sim$ 2.3 $k_{\rm B}$. See note 2 in SM.

\section*{Acknowledgements}
We thank Yu Jin, Elizabeth M.Y. Lee and Marco Govoni for useful discussions. This work was supported by MICCoM, as part of the Computational Materials Sciences Program funded by the U.S. Department of Energy, Office of Science, Basic Energy Sciences, Materials Sciences, and Engineering Division through Argonne National Laboratory. 
This work was also supported by the QNEXT hub (1F-60579). 
We acknowledge the computational resources at the University of Chicago's Research Computing Center and the Argonne Leadership Computing Facility.

\section*{Author contributions}
C.Z., F.G. and G.G. designed the calculations. 
C.Z. performed the calculations.  
All authors contributed to the
data analysis and the manuscript writing.

\section*{Competing interests}
The authors declare no competing interests.

\bibliography{Ref_main}
\bibliographystyle{apsrev4-2}

%TC:endignore

\end{document}

\title{Supplementary Information for: \\ Engineering the formation of spin-defects from first principles}

\author{Cunzhi Zhang}
\affiliation{Pritzker School of Molecular Engineering, University of Chicago, Chicago, IL
60637, USA}
\author{Francois Gygi}
\affiliation{Department of Computer Science, University of California Davis, Davis, CA 95616, USA}
\author {Giulia Galli}
\email{gagalli@uchicago.edu}
\affiliation{Pritzker School of Molecular Engineering, University of Chicago, Chicago, IL
60637, USA}
\affiliation{Department of Chemistry, University of Chicago, Chicago, IL 60637, USA}
\affiliation{Materials Science Division and Center for Molecular Engineering, Argonne National Laboratory, Lemont, IL 60439, USA}

\date{\today}

\maketitle
\clearpage

\section*{Note 1: Defect formation energies}
\begin{figure*}[hp!]
\centering
  \includegraphics[width=0.45\textwidth]{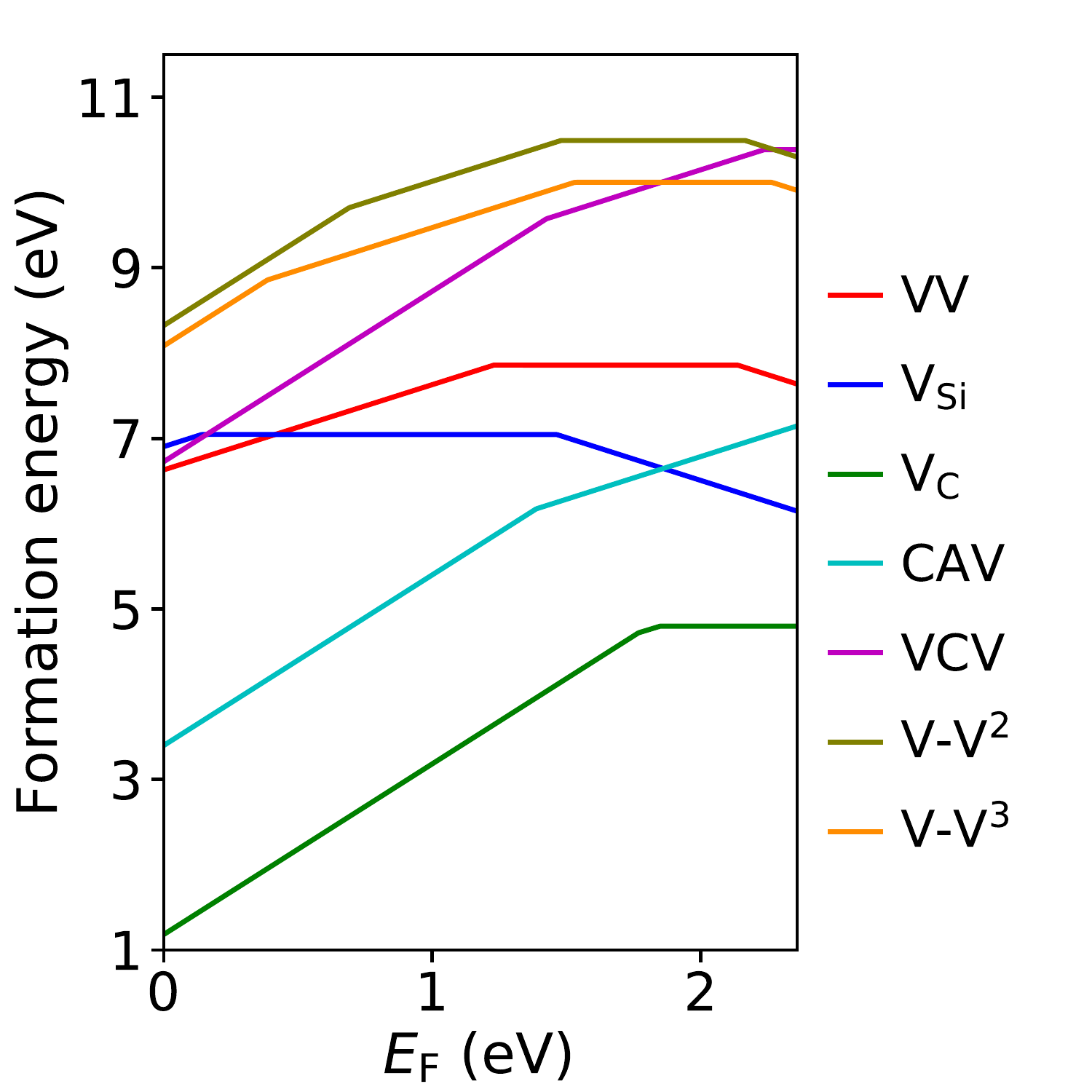}
  \caption{Formation energy of defects in 3C-SiC obtained under C-rich conditions, obtained using DFT and the DDH functional.  The geometry of these defects can be found in Fig. 1. See Methods in the manuscript.}
  \label{fig:ef_SM}
\end{figure*}

\clearpage
\section*{Note 2: Collective variables and free energy barriers}

\begin{figure*}[hp!]
\centering
  \includegraphics[width=0.9\textwidth]{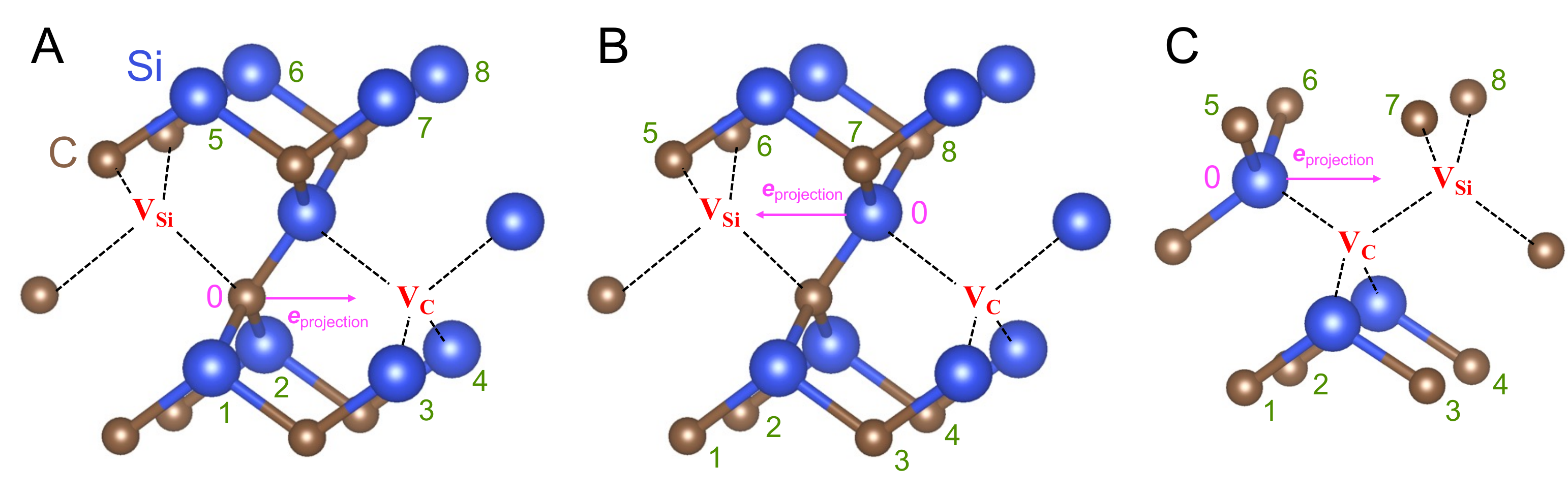}
  \caption{Collective variables (CV) for three paths investigated in our simulations. Number 0 denotes the moving atom; numbers 1-8 denote the gate atoms; projection vectors are also indicated. The three paths are: A: V-V$^3$ $\rightarrow$ VV @ V$\rm _C$; B: V-V$^3$ $\rightarrow$ VV @ V$_{\rm Si}$; C: first step in VV migration (see Fig. 1 in the manuscript).}
  \label{fig:CV_SM}
\end{figure*}
We obtained the free energy surface at 1,500 K using the adaptive biasing force method. In our enhanced sampling calculations, we considered three pathways, as shown in Fig. \ref{fig:CV_SM}, where we highlight the moving atom, gate atoms and projection vectors used to define the collective variable (see Methods in the manuscript). 
% We note that the performance {\bf ?} of the CV is not sensitive to the choice of gate atoms. 
For each pathway, we computed the free energy barrier $G_{\rm b}$ and entropy change $\Delta S$ from the initial to the transition state, which we denote  as forward (F) direction; we then computed the $G_{\rm b}$ and $\Delta S$ from the final to the transition state, which we denote as backward (B) direction. 
We summarize the computed values of $G_{\rm b}$ and $\Delta S$ in Table \ref{tab:barrier_SM}. 

For the three pathways studied here at 1,500 K, $G_{\rm b}$ decreases  by $\sim$ (0.11, 0.38) eV, relative to the value at 0 K. Using the harmonic approximation and classical statistics (see Note 4), we estimate $\Delta S$ in the range of $\sim$ (0.85, 2.90) $k_{\rm B}$. In particular, in going from the stable VV configuration to the transition state, the $G_{\rm b}$ decreases by $\sim$ 0.3 eV and the corresponding $\Delta S$ is $\sim$ 2.3 $k_{\rm B}$; the latter value is  used  to estimate the activation temperature (see Note 4).

We note that the values of $\Delta S$ were computed at the PBE level of theory ($\Delta S^{\rm PBE}$).
Using the harmonic approximation and classical statistics (see Note 4),
$\Delta S$ is determined from phonon frequencies. Assuming error cancellations between total energies computed at the PBE and DDH levels of theory, it is also reasonable to assume that $\Delta S^{\rm PBE} \sim \Delta S^{\rm DDH}$.

\begin{table}[hp!]
\centering
\caption{Free energy barriers $G_{\rm b}$ (eV),  free energy barrier change $\Delta G_{\rm b} = G_{\rm b} @\ 0\ {\rm K} - G_{\rm b} @\ 1,500\ {\rm K} $ 
(eV) and entropy change  $\Delta S = \Delta G_{\rm b} / T$ ($k_{\rm B}$), where $k_{\rm B}$ is the Boltzmann constant and $T$ = 1,500 K.}

\begin{tabular}{ |m{2.8cm}|m{2.cm}|m{2.cm}|m{2.cm}|m{2.cm}|m{3.5cm}| }
 \hline
 \multicolumn{6}{|c|}{  $G_{\rm b}$ } \\ \hline
 Pathways & 
 \multicolumn{2}{|l|}{ V-V$^3$ $\rightarrow$ VV @ V$\rm _C$ } 
 &
 \multicolumn{2}{|l|}{ V-V$^3$ $\rightarrow$ VV @ V$_{\rm Si}$ 
 } & 
 \multicolumn{1}{|l|}{ VV migration$^{\rm a}$ } \\ \hline
  Direction$^{\rm b}$ & F & B & F & B & F (=B) \\ \hline
  $G_{\rm b}$ @ 0 K & 2.35 & 4.34 & 1.29 & 3.28 & 3.07 \\  \hline
  $G_{\rm b}$ @ 1,500 K & 2.24 & 4.06 & 1.14 & 2.98 & 2.70 \\ \hline
  \multicolumn{6}{|c|}{  $\Delta G_{\rm b}$} \\ \hline
  $\Delta G_{\rm b}$ & 0.11 & 0.28 & 0.15 & 0.30 & 0.38 \\  \hline
  \multicolumn{6}{|c|}{  $\Delta S$ } \\  \hline
  $\Delta S$ & 0.85 & 2.17 & 1.16 & 2.32 & 2.90 \\ \hline
  
\end{tabular}

\begin{threeparttable}
\begin{tablenotes}\footnotesize
    \item[a] Only the first step in VV migration path is simulated, as shown in Fig. \ref{fig:CV_SM}C.
    \item[b] F refers to forward direction from the initial to the transition state; B refers to backward direction from the final to the transition state.
\end{tablenotes}
\end{threeparttable}

\label{tab:barrier_SM}
\end{table}

\clearpage

\section*{Note 3: Calculation of effective barriers}

During transformations occurring at high temperature ($T$), point defects may be in several charge ($q$) and spin ($s$) states different from the  thermodynamically stable ones.  Hence we should consider different energy barriers $E_{\rm b}(q, s)$. Moreover, the transition between different $q$ and $s$ states may occur at elevated $T$ due, e.g., to vibrational effects. For these reasons, we used effective barriers $E_{\rm b,\ EFF}$\cite{gerst2004PRB_EFF,bruneval2011PRB_EFF} to describe atomic processes occurring at high $T$, instead of simple barriers $E_{\rm b}$.  

% We found that the  energy splitting  between different spin states is usually small relative to the  energy barriers $\sim$ several eV.
% ({\bf small relative to what?}).

In our NEB calculations, we first  determined the most stable $s$ state for each image at a given $q$; the corresponding total energies and atomic forces were then used to determine the minimum energy path and $E_{\rm b}$. Hence the final
$E_{\rm b}$ obtained in this way is only a function of $q$, with the effect of the $s$ degree of freedom (DOF) included implicitly.
Our treatment of the spin DOFs assumes an instantaneous equilibration of spin states during defects' transformations.
Although we could not estimate the timescale of $s$ transitions via spin-orbital-coupling or spin-phonon interactions at $\sim$ 1,000 K, we found that 
in general considering different spin states affects only slightly the computed $E_{\rm b}$ (see Methods in the manuscript).
 
We computed $E_{\rm b,\ EFF}$ based on $E_{\rm b}$ and defect formation energies, considering only the charge DOF (see Computational strategy in the manuscript). We assumed the charge state $q$ to be  preserved during defect transformations, due to the short lifetime of barrier crossing over the transition state, and we considered $q$ transitions  at the initial and final states of a given path.

In the case of the dissociation of complex defects, involving multiple steps, $E_{\rm b,\ EFF}$ was estimated from the  binding energy and diffusion barriers. For instance, $E_{\rm b,\ EFF}$ for the CAV dissociation process was obtained as the sum of the  binding energy ( C$_{\rm Si}$ \& V$_{\rm C}$) and the  $E_{\rm b,\ EFF}$ of V$_{\rm C}$ migration.

% we assumed the $q$ to be  preserved during defect transformations.{\bf ?}
% \begin{equation}
%    E_{\rm b,\ EFF} (E_{\rm F}) = \underset{q \in \{ 2, 1, 0, -1, -2 \} }{\text{min}}
%    \big\{ \Delta E_{\rm f}(q, E_{\rm F})
%    + E_{\rm b}(q) \big\}
% \end{equation}
% where $\Delta E_{\rm f}(q, E_{\rm F})$ is formation energy difference for a given $q$, with respect to the most stable one at given $E_{\rm F}$; $E_{\rm b}(q)$ is barrier at $q$ obtained from NEB. 
For $E_{\rm b,\ EFF}$ to be accurate, the transition between different $q$ states needs to be fast 
relative to the transformation of defects into different configurations.
We estimated the $q$ transition, via carrier capture or emission, is indeed fast at high $T$, as indicated below. 
% We also note that for processes with low $E_{\rm b}$, e.g. interstitial jump, the defect transformations can be fast and comparable to the $q$ or $s$ transition; in these cases, the use of $E_{\rm b,\ EFF}$ can be problematic.   {\bf ?} 

Under equilibrium condition, the $q$ transition rate ($g$)~\cite{bruneval2011PRB_EFF,bourgoin1983_Q} can be estimated as:  
\begin{equation}
   g(T) = \sigma \langle v \rangle N \gamma 
   \exp( - \frac{ \Delta E }{ k_{\rm B}T } )
\end{equation}
where $\sigma$ is the capture cross section; $\langle v \rangle$ is the average thermal velocity of carriers; $N$ is the effective density of states (DOS); $\gamma$ is the degeneracy factor; $\Delta E$ is the energy difference between defect levels and the closest band edge; $k_{\rm B}$ is the Boltzmann constant.

Given $N(T)$ $\sim T^{3/2}$ \cite{kimoto2014book_SiC, neamen2012_fourth}, we estimated $N (1,000\ \rm K)$ $\sim$ $N (300\ \rm K) \times (10/3)^{3/2}$. $N (300\ \rm K)$ was measured experimentally \cite{kimoto2014book_SiC}. 
Given $\langle v \rangle (T) $ $\sim$ $(3 k_{\rm B}T / m_{0})^{0.5}$, where $m_{0}$ is the mass of stationary electron; we estimated $\langle v \rangle (1,000\ \rm K)$ $\sim$ $2 \times 10^7$ cm/s. Based on experiments, for most of deep levels in SiC, $\sigma$ are in the range of (10$^{-17}$, 10$^{-14}$) cm$^2$ \cite{nakane2021JAP_cross,hazdra2019PSS_cross,danno2007JAP_cross,tunhuma2018JAP_cross}.
We assumed $\gamma$ = 1 and $\Delta E$ $\sim$ 1 eV, approximately  half the band gap of 3C-SiC. Then, we obtained the timescale, $1/g(1,000\ \rm K)$, in the range of $\sim$ (10$^{-8}$, 10$^{-5}$) s. For comparison, for defect transformations at 1,000 K with barrier $\sim$ 3 eV, the timescale is estimated to be $\sim$ 10 s.

\clearpage
\section*{Note 4: Calculations of the activation temperature}

According to the harmonic transition state theory\cite{viney1957JPCS_HTST,li2007MRS_HTST}, the jump frequency $\Gamma$ can be calculated as:
\begin{equation}
  \Gamma = \Gamma_{0} \exp( - G_{\rm b} / k_{\rm B}T )
\end{equation} 
where $\Gamma_{0}$ is the attempt frequency and  $G_{\rm b}$ the free energy barrier of a given process. 

During defect transformations, we assumed the system volume to be constant, hence:
\begin{equation}
  G_{\rm b} = \Delta U - T\Delta S
\end{equation} 
where $\Delta U$ is the change in internal energy from the initial to the transition state of a given path; $\Delta S$ is the change in entropy from the initial to the transition state of the path.  Note that we computed $\Delta S$ for three paths only, due to the computational cost,  and found values varying within $\sim$ (0.85, 2.90) $k_{\rm B}$ (see Note 2).  Based on the harmonic approximation and classical statistics, the change in kinetic energy is 0 eV, and the change in potential energy is $E_{\rm b}$ (barrier at 0 K), enforced by the equi-partition theorem. In addition, $\Delta S$ is constant, as determined by calculations of phonon frequencies\cite{viney1957JPCS_HTST,li2007MRS_HTST}. Therefore, we obtain:
\begin{equation}
\label{eq:gamma_SM}
  \Gamma \approx \Gamma_{0} \exp( \frac{\Delta S}{k_{\rm B}} )
  \exp( - E_{\rm b} / k_{\rm B}T )
\end{equation} 

The activation temperature $T_{a}$ is defined as the $T$ above which a process is thermally activated. Based on Eq. \ref{eq:gamma_SM}, $T_{a}$ can be written as: 
\begin{equation}
\label{eq:ta_SM}
  T_{\rm a} = \Big[ k_{\rm B} 
  \ln( \Gamma_{0} \exp( \frac{\Delta S}{k_{\rm B}}) 
  / \Gamma ) \Big]^{-1} 
  \times E_{\rm b,\ EFF} 
%   \approx 331 \times E_{\rm b}\ ({\rm K})
\end{equation} 

Similar to previous studies, we approximated $\Gamma_{0}$ to be 1.6 $\times$ 10$^{13}$ Hz \cite{rauls2003PRB_TS}; jump frequency $\Gamma$ to be 0.1 Hz \cite{kyrts2017PRB_1Hz,froda2021PRM_1min}; $\Delta S$ to be 2.3 $k_{\rm B}$. This value corresponds to our estimate of  the $\Delta S$ from the stable VV configuration to the transition state (see Note 2).
We obtained a prefactor (inverse of the quantity within square brackets in Eq. \ref{eq:ta_SM} ) of 331 K/eV. 
% We also noted that anharmonic effects are partially included into $\Delta S$, as phonon renormalization at 1500 K is captured by enhanced sampling MD. 
A simple sensitivity analysis shows that such prefactor is relatively insensitive to the choice of $\Gamma$ and $\Delta S$. For example, by varying $\Gamma$ in the range of (0.01, 1) Hz (with all other parameters fixed), the prefactor changes by $<$ $\sim$ 7 \%; by varying $\Delta S$ in the range of (0.85, 2.90) $k_{\rm B}$, the prefactor changes by $<$ $\sim$ 4 \%. 

We systematically investigated the thermal expansion and entropic effects on computed energy barriers. We found that considering lattice expansion at 1,500 K leads to a minor change  of $E_{\rm b}$ (which is on the order of several eV) of approximately  $\sim \pm$ 0.1 eV for most processes,  with the exception of  small carbon-clusters' formation for which  differences in energy barriers are $\sim \pm$ 0.3 eV. 
We found that due to entropic effects, our computed free energy barriers are lowered by $\sim$ (0.11, 0.38) eV at 1,500 K, relative to those obtained at 0 K (see Note 2), consistent  with estimates based on the harmonic approximation~\cite{rauls2003PRB_TS}. 
As a result, we estimate that the variation of $T_{\rm a}$ due to thermal expansion and entropic effects is less than 10 $\%$ .

\clearpage
\section*{Note 5: Calculations of the Fermi level}

After irradiation or implantation of SiC samples, multiple defects can be created including interstitials, antisites, substitutionals and vacancies. 
% Given that: 1) interstitials are annealed out at low $T$; 2) antisites and substitutions are mostly immobile, only vacancies are relevant at high $T$ annealing for VV creation processes.  
% We note that antisites in SiC are generally neutral within the band gap\cite{yan2020JAP_VSi}, while substitutional atoms can be charged. 
In order to obtain an accurate value of $E_{\rm F}$ we need to consider both external doping and the charge state of the defects created in the sample. 

\begin{figure*}[!htp]
\centering
  \includegraphics[width=0.99\textwidth]{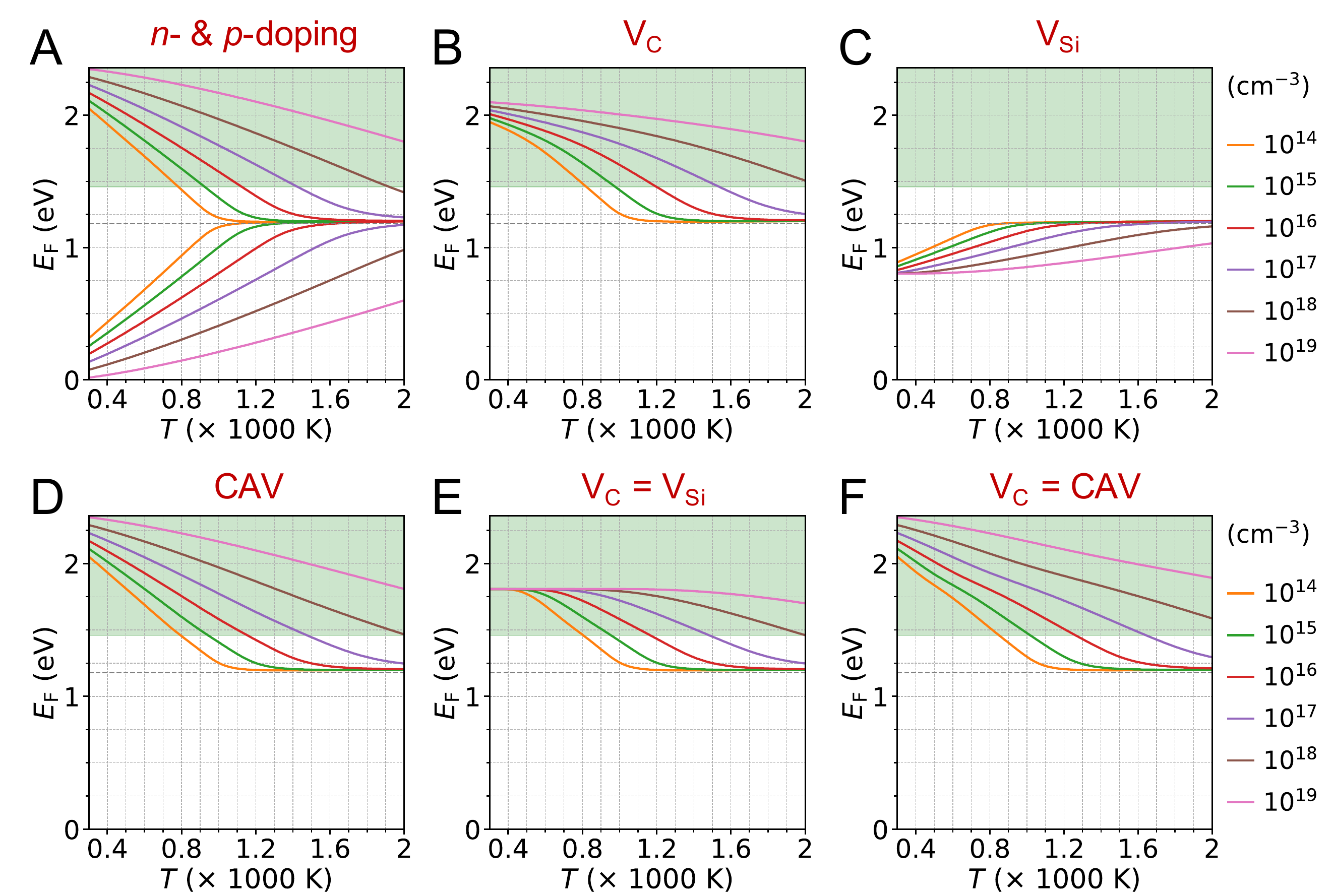}
  \caption{Fermi level as a function of $T$ in 3C-SiC. The Fermi level is referred to the top of the valence band. We consider the doping or defects density in the range of (10$^{14}$, 10$^{19}$) cm$^{-3}$. A: presence of $n$- or $p$-doping only. B: presence of carbon vacancy (V$_{\rm C}$) only. C: presence of silicon vacancy (V$_{\rm Si}$) only. D: presence of carbon antisite vacancy  (CAV) only. E: presence of V$_{\rm C}$ and V$_{\rm Si}$ of the same amount. F: presence of V$_{\rm C}$ and CAV of the same amount. The value of mid-gap for 3C-SiC is indicated by a grey dashed line; the green-region for $E_{\rm F}$ $>$ 1.46 eV indicates the suitable conditions for the VV creation.}
  \label{fig:fermi_SM}
\end{figure*}

\begin{figure*}[!htp]
\centering
  \includegraphics[width=0.5\textwidth]{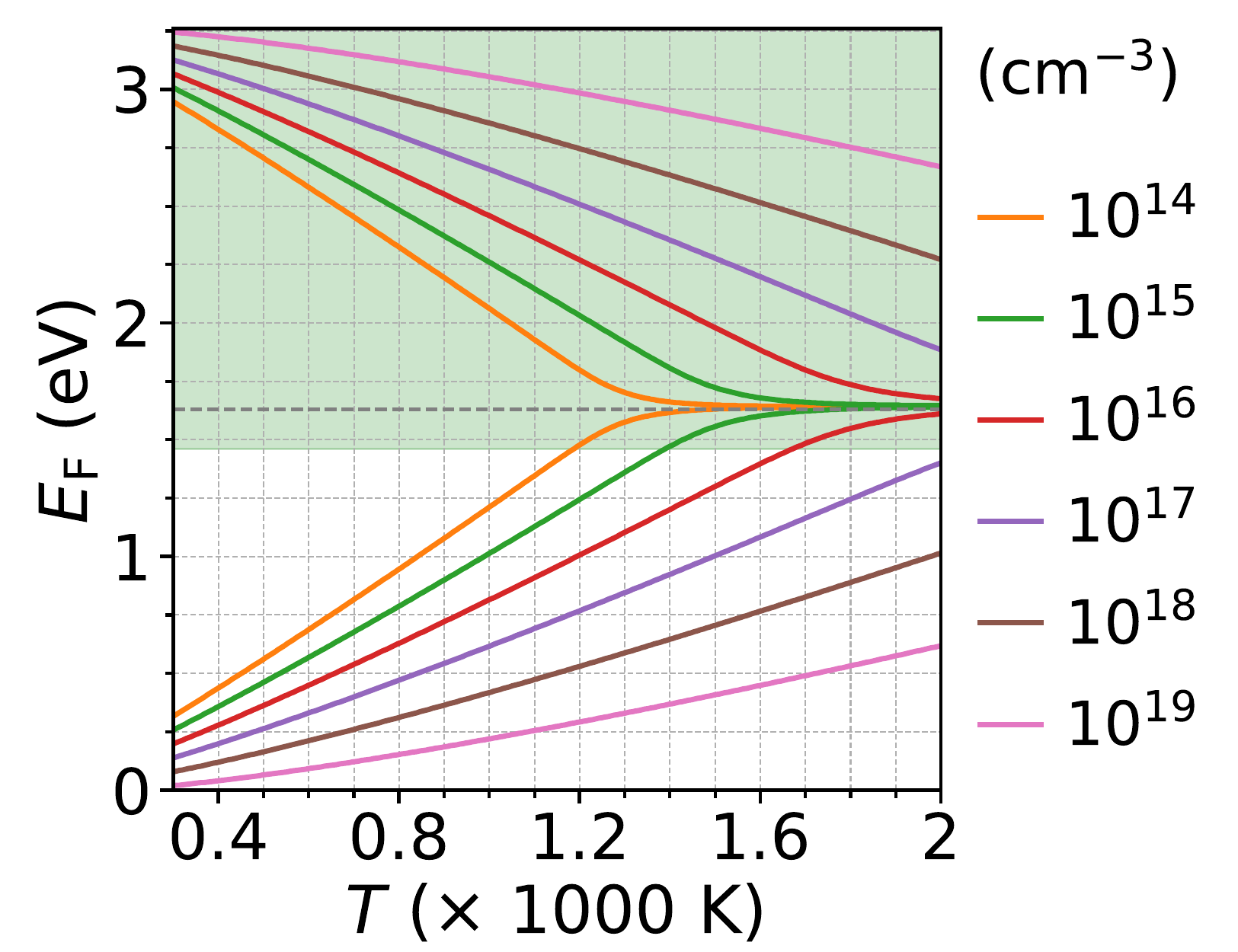}
  \caption{Fermi level as a function of $T$ in 4H-SiC.
  The Fermi level is referred to the top of the valence band. We consider the $n$- or $p$-doping density in the range of (10$^{14}$, 10$^{19}$) cm$^{-3}$. The value of mid-gap for 4H-SiC is indicated by a grey dashed line; the green-region for $E_{\rm F}$ $>$ 1.46 eV indicates the suitable conditions for the VV creation.
  }
  \label{fig:fermi_4H_SM}
\end{figure*}

In this study, we took into account several vacancies: V$_{\rm C}$, V$_{\rm Si}$ and CAV, which are relevant to the VV creation processes. We determined $E_{\rm F}$ using the following equations \cite{neamen2012_fourth,ma2011PRB_compensation,bucker2019CPC_compensation}:
\begin{equation}
\label{eq:ef_SM}
  \begin{cases}
    a) & n_{0}p_{0} = N_{\rm C} N_{\rm V} 
      \exp( - E_{\rm g} / k_{\rm B}T ) \\
    b) & n_{0} + N_{\rm a} = p_{0} + N_{\rm d} + \sum_{ {\rm X \in (V_C, V_{Si}, CAV) } }
         \sum_{q} q \times N( {\rm X} ^ {q} )\\
    % b) & n_{0} + N({\rm V_{Si}^{\ -}}) + N_{\rm a} = p_{0} +
    % N({ \rm CAV^{+1} }) + 2 \times N({ \rm V_{C}^{\ +2} }) + 
    % N({ \rm V_{C}^{\ +1} }) + N_{\rm d}\ \text{(e.g.)}\\
    c) & N( {\rm X} ^ {q} ) \propto N_{\rm X} \textsl{g}_{ {\rm X} ^ {q} } 
    \exp( - E_{\rm f} ({\rm X}^{q}) / k_{\rm B}T ) \\
    d) & E_{\rm F} - E_{\rm V} = k_{\rm B}T 
    \ln(N_{\rm V} / p_{0})
  \end{cases}   
\end{equation} 
% $N_{\rm V}=2 (\frac{2\pi m_{p} k_{\rm B} T}{ h^{2} }) ^ {3/2}$
where $n_0$ is the electron density; $p_0$ is the hole density; $N_{\rm C}$ is the effective conduction band DOS; $N_{\rm V}$ is the effective valence band DOS; $E_{\rm g}$ is the band gap; $N_{\rm a}$ is the acceptor density; $N_{\rm d}$ is the donor density; X$^q$ stands for defect X in charge state $q$; $N( {\rm X}^q )$ is the density of X$^q$; $N_{\rm X}$ is the total density of defect X; $\textsl{g}$ is the degeneracy factor; taken as 1; $E_{\rm f}$ is the defect formation energy (see Note 1); $E_{\rm V}$ is the valence band maximum (VBM) energy. 

Eq. \ref{eq:ef_SM}-$b$ expresses the charge neutrality condition, incorporating the effects of defects. These equations need to be solved self-consistently. 
Here, we performed a line-search by step-wisely increasing $E_{\rm F}$ from VBM to the conduction band minimum; we determined $E_{\rm F}$ as the value, which makes Eq. \ref{eq:ef_SM}-$b$ satisfied with a minimal error.
The electronic properties of SiC were obtained from the Appendix C of the book \cite{kimoto2014book_SiC}. For simplicity, we ignored the $T$ dependence of these parameters: 1) $E_{\rm g}$ at 300 K was used; 2) $E_{\rm f}$ diagram at 0 K was used (Fig. \ref{fig:ef_SM}). We deduced the DOS effective masses based on the measured $N_{\rm C}$ and $N_{\rm V}$ at 300 K, which were then used to compute $N_{\rm C}$ and $N_{\rm V}$ at various $T$. 
Some of our results are shown in Fig. \ref{fig:fermi_SM} and \ref{fig:fermi_4H_SM}. We note our $E_{\rm F}$ may be over-estimated, since band gap decreases at high $T$.

Overall, the calculation of $E_{\rm F}$ here should be taken as a qualitative estimate, as we: 1) ignored the $T$ dependence of electronic properties, e.g. band gap of SiC; 2) ignored the effects of other defects, in addition to V$_{\rm C}$, V$_{\rm Si}$ and CAV. More accurate treatment is beyond the scope of this work. 

% \bibliography{Ref_main,Ref_SM}
\bibliography{Ref_SM}

\bibliographystyle{apsrev4-2}